\documentclass[12pt, letterpaper]{article}
\usepackage{graphicx}
\usepackage{amsmath}
\usepackage{amssymb}
\usepackage{url} 
\usepackage{pdflscape}

\title{Interaction of Space Weather Phenomena With Mars Plasma Environment During Solar Minimum 23/24}
\author{P. Kajdi\v{c}$^1$, B. S\'anchez-Cano$^2$, L. Neves-Ribeiro$^{3,4}$, O. Witasse$^5$, G. C. Bernal$^3$,\\ D. Rojas-Castillo$^{1,6}$, H. Nilsson$^7$, A. Fedorov$^8$}

\begin{document}

\maketitle

 \begin{center}
$^1$Instituto de Geof\' isica, Universidad Nacional Aut\'onoma de M\'exico, Mexico City, Mexico,
$^2$University of Leicester, Department of Physics and Astronomy, School of Physics and Astronomy, Leicester, United Kingdom,
$^3$Instituto de Matem\'atica, Estat\' istica e F\' isica, Universidade Federal do Rio Grande, Rio Grande do Sul, Brasil,
$^4$Departamento de F\' isica de Plasmas y de Interacci\'on de Radiaci\'on con la Materia, Instituto de Ciencias Nucleares, Universidad Nacional Aut\'onoma de M\'exico,
$^5$European Space Agency (ESA), European Space Research and Technology Centre (ESTEC), Keplerlaan 1, 2201 AZ Noordwijk, The Netherlands,
$^6$Space Research Institute, Austrian Academy of Sciences, Graz, Austria,
$^7$Institutet f\"or Rymdfysik, Kiruna, Sweden,
$^8$Institut de Recherche en Astrophysique et Plan\'tologie, 9, Avenue du Colonel ROCHE, BP 4346, 31028 Toulouse Cedex 4, France
\end{center}

{\bf Keypoints}
\begin{enumerate}
\item Interaction of Mars' plasma environment with the solar wind structures during solar minimum 23/24.
\item Moderate solar wind structures interacting with Mars may cause enhanced transterminator planetary ion flows.
\item During such perturbed times the planetary ion transterminator flow exhibits strong variations of relative ion abundances.
\end{enumerate}

\section*{Abstract}
We study the interaction of three solar wind structures, two stream interaction regions and one interplanetary coronal mass ejection, with Mars' plasma environment during 20-27 November 2007. This period corresponds to the solar minimum between the solar cycles 23 and 24 which was characterized by very low values of the solar wind density and dynamic pressure and low IMF magnitude. During that time the Mars-Express orbit was in the terminator plane, while the Earth, Sun, and Mars were almost aligned, so we use the ACE and STEREO probes as solar wind monitors in order to identify and characterize the structures that later hit Mars. We find that the passage of these structures caused strong variations of in the bow shock location (between 2.2 and 3.0~R$_M$), compression of the magnetospheric cavity (up to 45~\%) and an increased transterminator flow below 2~R$_M$ (by a factor of $\leq$8). This study shows that during times of low solar activity, modest space weather phenomena may cause large variations of plasma flow at Mars.

\section{Introduction}
\label{sec:introduction}
``
Mars currently does not possess a global magnetic field and only remnant crustal fields exist on its surface, mostly on the southern hemisphere (\cite{acuna:1998}). Due to this, the solar wind (SW) interacts directly with the Martian upper atmosphere and ionosphere, which leads to significant loss of the planet's atmospheric material.

In the absence of a global magnetosphere, it is the Martian ionosphere which represents an obstacle for the SW and the interplanetary magnetic field (IMF). The super alfv\'enic and supermagnetosonic SW cannot penetrate into it but it is forced to decelerate and flow around the ionosphere. Part of the deceleration occurs at the bow shock (BS) that stands upstream of Mars and was first detected by Mariner-IV, Mars 2,3,5 and Phobos-2 (\cite{sagdeev:1989}) spacecraft at a subsolar distance of $\sim$3 Martian radii (R$_M$).

As the SW slows down, the IMF drapes around and piles up upstream of the ionosphere, forming an induced magnetosphere. The region where the IMF piles up is called the magnetic pile-up region (MPR) and is delimited by a magnetic pile-up boundary (MPB).
Inside the MPR the magnetic field rotates, its magnitude increases, fluctuations are reduced and energetic particle fluxes diminish. The volume of space delimited by the MPB is also called the magnetospheric cavity (\cite{dubinin:2009}).

The region between the MPB and the BS is called the {\it magnetosheath} and is characterized by turbulent B-fields and energized electrons (\cite{acuna:1998}). The planet's ionosphere, MPR and the magnetosheath form what is known as near-Mars plasma environment.

Normally the hot magnetosheath plasma and the planet's ionospheric plasma do not mix. They are separated by boundaries that were defined based on observational data. {\it The ionopause} was first inferred from radio occultation measurements (\cite{kliore:1992}) and confirmed by \cite{acuna:1998}). It is located at altitudes as low as 200-300~km (see also \cite{duru:2009, vogt:2015}) and corresponds to the region where energetic (E$>$50~eV) electron fluxes drop by an order of magnitude, while intense cold electron fluxes (E$<$10~eV) are observed (\cite{acuna:1998}). The {\it photoelectron boundary} (PEB, i.e., \cite{mitchell:2001, lundin:2004, garnier:2017}) separates ionospheric electrons from shocked magnetosheath electrons at higher altitudes. Photoelectrons (energy range 20-50~eV) at Mars are produced due to extreme ultraviolet (EUV) light and X-rays emitted by the Sun. Especially important is the 30.4~nm Helium II spectral line that photoionizes neutral CO$_2$ and O particles. Finally,  due to the lack of magnetometers onboard the Mars-Express (MEX, \cite{schmidt:2003}) mission, the {\it induced magnetosphere boundary} (IMB) was defined as the region where the SW stops while its interior is dominated by plasma of planetary origin (\cite{lundin:2004}).

\begin{figure}
\centering
\includegraphics[width=1.0\textwidth]{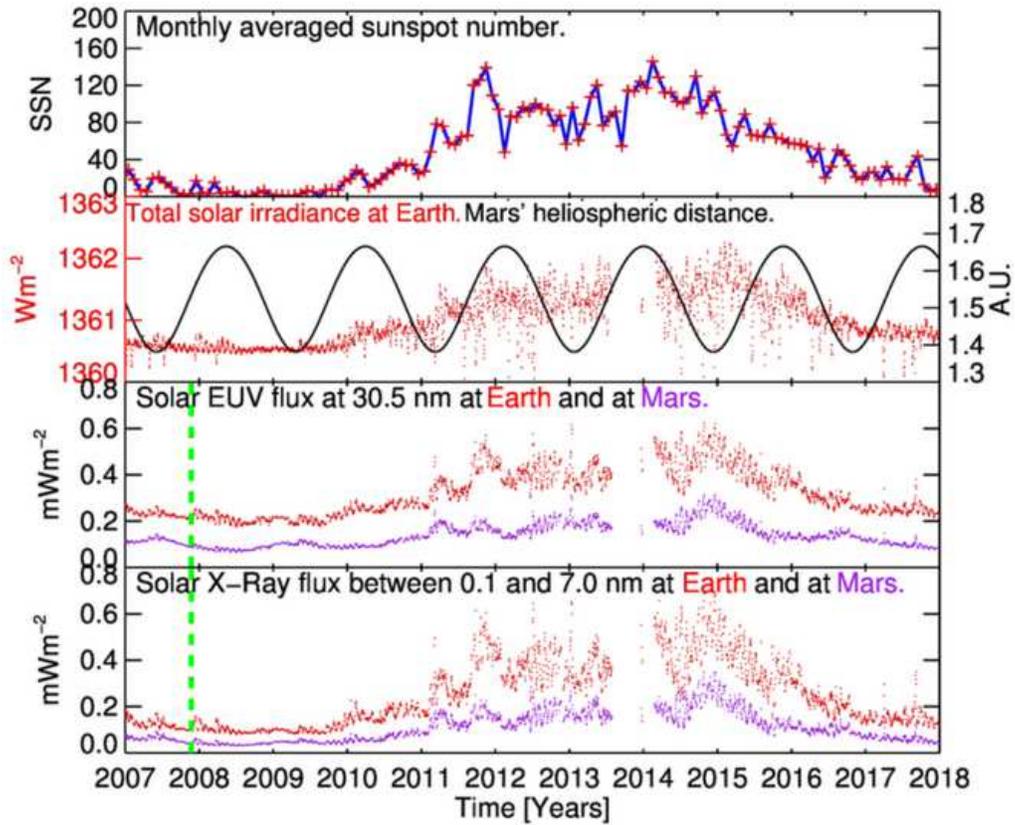}
\caption{From top to bottom: a) monthly averaged sunspot number, b) total solar irradiance at Earth (red dots) and Mars' heliocentric distance (black curve), c) solar EUV and d) solar X-ray fluxes at Earth (red) and Mars (purple).}
\label{fig:fluxes}
 \end{figure}

There are many phenomena that influence the Martian ionosphere/induced magnetosphere system. Among them are the varying EUV and X-ray fluxes during solar cycles and during solar flares (see Figure~\ref{fig:fluxes} and \cite{fletcher:2011}), varying SW dynamic pressure (P$_{dyn,SW}$), transient structures such as  stream interaction regions (SIR) including corotating interaction regions  (CIR , i.e. \cite{gosling:1999})  and interplanetary coronal mass ejections (ICME, \cite{sheeley:1985}), solar energetic particle (SEP, i.e. \cite{reames:1996, schwenn:2006}) and energetic storm particles (ESP, \cite{cohen:2006}) events.
We briefly describe the effects of some of these events, relevant to the present study, on the Mars' plasma environment.

\subsection{Stream Interaction Regions}
The interaction of the Mars plasma environment with SIRs has been studied by \cite{dubinin:2009}, \cite{morgan:2010}, \cite{opgenoorth:2013}, \cite{lee:2017}, \cite{lee:2018} and others. Strong perturbations of the Martian magnetosphere and ionosphere during SIR passages were reported together with increased ionospheric electron densities. In some cases the magnetic barrier ceased to be a shield for the incoming SW and large amount of SW could penetrate into the magnetosphere sweeping out cold ionospheric plasma. A statistical studies of CIRs on the atmosphere of Mars was performed by \cite{edberg:2010a} and by \cite{nilsson:2011} who reported that ion escape fluxes incremented by a factor of $\sim$2.5 during pressure pulses produced by CIRs.

\subsection{SW dynamic pressure}
Another factor that changes during solar cycle is the SW dynamic pressure P$_{dyn,SW}$. P$_{dyn,SW}$ increases may occur, for example, behind interplanetary (IP) shocks, during fast SW streams or inside SIRs. \cite{harada:2017}, for example, studied the response of the Martian ionosphere to a SIR related IP shock. A sharp increase in the local ionospheric B-field at 478 km at a solar zenith angle of 78 degrees was observed, which points towards the compression of the Martian induced magnetosphere.

The effect of the high P$_{dyn,SW}$ on the Martian plasma environment was studied by several authors who obtained some contradicting results. \cite{edberg:2009} showed that during high P$_{dyn,SW}$ the SW can penetrate into the ionosphere and enhance its erosion. \cite{nilsson:2011} also found a correlation between ion escape and the P$_{dyn,SW}$. \cite{ramstad:2017} on the other hand found a weak inverse dependence of the ion escape rate on the P$_{dyn,SW}$.

\subsection{Interplanetary Coronal Mass Ejections}
The interaction of the Mars plasma system with interplanetary coronal mass ejections has been a topic of several works in the recent past. Most of these works study impacts of large ICMEs, such as the Halloween events (\cite{crider:2005}, \cite{espley:2005}), the 14 October 2014 (\cite{witasse:2017}), the 5 December 2005 (\cite{morgan:2014}), 8 March 2015 (\cite{jakosky:2015, curry:2015, ma:2017, duru:2017}) and 13 September 2017 (\cite{guo:2017, lee:2018, ramstad:2018, xu:2018, ma:2018, sanchezcano:2019}) events.

Large ICMEs are often accompanied by intense SEP/ESP events and solar flares. It is thought that SEP/ESP events augment ion loss at unmagnetized planets through extra charging of the ionosphere (\cite{futuaana:2008, opgenoorth:2013,morgan:2014}). SEP particles are able to penetrate the induced magnetic barrier and precipitate into the planetary atmospheres, modifying heating, dissociation, and ionization rates and causing auroral emissions (\cite{ramstad:2018}).

The compression of the induced magnetosphere  reduces the altitude of the MPB (from the typical value of 800-1200 km to only 400 km, see \cite{crider:2005}) thereby exposing the plasma from the ionosphere and lower exosphere directly to the SW flow. This enables charge exchange between the SW ions and the atmospheric neutrals which leads to further enhancement of the ionospheric ion escape. \cite{ehresmann:2018} studied effect the energetic particles on Mars' surface during the September 2017 event. It was found that fluxes of protons with energies below 100~MeV were enhanced by  a factor of 30 and those with higher energies by a factor of four. A completely different result however was obtain by \cite{ramstad:2017b} who studied an impact of an extremely strong ICME that hit Mars on 12 July 2011. These authors found no significant increase of the ion escape rates during that time.

Numerical simulations of such events (e.g., \cite{jakosky:2015, dong:2014, curry:2015, ma:2017, romanelli:2018}) predict planetary ion escape rates increase by a factor of 2--10. \cite{curry:2015} proposed that highest O$^+$ escape rates occur during the passage of the sheath region, which is when the P$_{dyn, SW}$ is largest.

Other effects of large ICMEs include oscillations of the ionosphere, appearance of ionization at low altitude in the atmosphere (\cite{sanchezcano:2019}) and extension of the ionospheric peak to up to 115$^\circ$ of solar zenith angle (SZA) (as opposed to $\leq$100$^\circ$ during quiet times, \cite{morgan:2014}) and variations of all plasma boundaries (\cite{trotignon:2006, edberg:2010, xu:2018, ma:2018, romanelli:2018}).

\subsection{Small events during solar minima}
 To our best knowledge there has been only one study that dealt with the interaction between Mars plasma environment and small scale ICME (lasting only about 5~hours) and SIR during solar minimum (\cite{sanchezcano:2017}).

\cite{sanchezcano:2017} studied a Mars plasma system response to two SW disturbances that arrived to Mars within $\sim$1~day on 7-8 March 2008. The first disturbance was a small scale ICME-like structure and the second a high speed solar wind.
The structure caused a strong compression of the Martian induced magnetosphere and ionosphere lasting for $\sim$3 MEX orbits ($\sim$20~hours). The fast SW stream caused yet another compression of both regions and fragmentation of the ionosphere for several days. It was concluded by the authors that during solar minima, small SW transients may cause large and long lasting perturbations of the Mars' plasma system.

In this work we study the impact of SW transients on Martian plasma environment during solar minimum 23/24. Specifically, we perform a case study of impact of an ICME and two SIRs on the Mars plasma environment starting on 22 November 2007. Although the properties of these transients were similar to those studied by \cite{sanchezcano:2017}, there are some important differences between the two events. In our case MEX orbit lies in the terminator plane.  This enables us to directly study ion flow across the terminator plane. Also, in our case the SIRs and the ICME arrive separately so impacts of each transient can be studied separately.

With this study we aim to contribute to the understanding of what is the impact of relatively small SW transients on the Mars' plasma environment during times of low solar activity. We mainly focus on how the plasma boundaries are affected by these phenomena and how the transterminator flow behaves during such perturbed times. Some of the plasma boundaries, such as the MPB, are partially idenitified without magnetic field measurements. Some boundaries such as PEB, IMB, MPB and ionopause may lie close togethermaking it difficult to clearly differentiate among them, though overall their presence  indicate the crossing from the magnetosheath into the ionospheric environment. Hence in the rest of the text, whenever we observe the spacecraft crossing from the magnetosheath into the ionosphere, we talk about the crossing of the IMB.

\section{Instrumentation}
\label{sec:instrumentation}
During the studied time period, the Sun, Earth and Mars were aligned in that same order. This enables us to use the two Solar Terrestrial Relations Observatory (STEREO) probes, which were located close to Earth at the time, and the Advance Composition Explorer (ACE) as SW monitors in order to understand better the observations of the Mars Express (MEX) mission.

STEREO mission consists of two identical probes in the orbit around the Sun at approximately 1~AU. We use data from several instruments from the In situ Measurements of Particles And CME Transients (IMPACT, \cite{luhmann:2008}) suite:  (i)  magnetometer (MAG, \cite{acuna:2008}) that provides accurate measurements of the magnetic field with subsecond time resolution and (ii) Plasma and Suprathermal Ion Composition (PLASTIC, \cite{galvin:2008})  that provides ion measurement in the 0.3-80~keV energy range.

ACE mission is located at the Lagrangian L1 point in front of the Earth. Its magnetic field data were obtained by the magnetic field experiment (\cite{smith:1998}) and the bulk SW moments were obtained by the Solar Wind Electron Proton Alpha Monitor (SWEPAM, \cite{mccomas:1998}).

For measurements at Mars we use the data from the Analyzer of Space Plasmas and Energetic Atoms (ASPERA-3, \cite{barabash:2004, barabash:2006}) experiment onboard MEX. ASPERA-3 has several sensors among which are the Electron Spectrometer (ELS) and the Ion Mass Analyzer (IMA). ELS covers the range between 0.01-20 keV with a field of view of 4$^\circ\times$360$^\circ$ (\cite{franz:2010}).
IMA consists of an electrostatic deflector which is followed by a top hat electrostatic analyzer, a circular magnetic separator and finally by a circular microchannel plate (MCP). IMA had several software patches installed throughout time. These patches allowed for an expansion of the instrument's energy range in which it operates (e.g. \cite{lundin:2008a, franz:2010, nilsson:2010, rojascastillo:2018}). In November 2007 the IMA energy range was 0.01-36~keV/q and the energy resolution was of 7~\% with an instantaneous field of view of 4.5$^\circ\times$360$^\circ$. An elevation scanning of $\pm$45$^\circ$ for energues $<$50~eV allows IMA to obtain a 3D spectrum at these energies after 192~s when operating in the maximum operational mode.

It should be noticed that for ions below 50~eV there is no elevation scanning and the IMA field of view is therefore limited to 6$^\circ$×360$^\circ$ (\cite{franz:2010}).
This means that the number density of the low energy ions could be underestimated.

For this work we use the IMA plasma data sets that are available through AMDA web tool (http://amda.irap.omp.eu/) and have been processed using similar as that described in \cite{fedorov:2011} for very similar IMA instrument onboard Venus Express in order to reconstruct the ion distributions and the plasma moments.
We use IMA measurements for O$^{+}$ and O$_2^+$ as they are the main heavy ion species in the Mars environment and are well resolved by the instrument (\cite{rojascastillo:2018}).

We also use the High-Energy Neutron Detector (HEND) on board the Mars Odyssey that is composed of two scintillators that provide measurements of charged particles and energetic photons (e.g. \cite{boynton:2004, zeitlin:2010, SanchezCano:2018}). We use the data on particles with energies between $\sim$195~keV - 1000~keV (\cite{livshits:2006}) provided by the outer scintillator.

\section{Observations at 1 A.U.}

\subsection{EUV and X-ray data}
\label{subsec:euvxray}
In order to put the observed event in the context of conditions prevailing in the interplanetary (IP) space during the time of interest, we show in Figure~\ref{fig:fluxes}a) the monthly smoothed sunspot number (from Sunspot Index and Long-term Solar Observations available at http://www.sidc.be/silso/), b) the total solar irradiance at Earth (red dots, from SOlar Radiation $\&$ Climate Experiment (SORCE), available at http://lasp.colorado.edu/home/sorce/) together with the Mars' heliocentric distance (black line), c) the solar EUV flux at 30.5~nm at Earth (red) and Mars (purple) and d) the solar X-ray flux between 0.1~nm and 7.0~nm at Earth (red) and scaled (as $1/r^2$) to Mars (purple) between 1 January 2007 - 31 December 2017. The green dashed line on the bottom panel marks the period when the event studied here was observed. It can be seen that the period of interest was characterized by very low EUV and X-ray fluxes compared to the rest of the solar cycle.

\subsection{IMF and plasma data}
\label{subsec:imfplasma}
Next we use the data from STEREO A (STA), STEREO B (STB) and ACE missions in order to describe the state of the SW and the interplanetary magnetic field (IMF) between 17-24 November 2007. During this time period the three spacecraft observed SW structures that later propagated to Mars (see also Table~\ref{tab:times}).

\begin{figure}[h]
 \centering
 \includegraphics[width=0.8\textwidth]{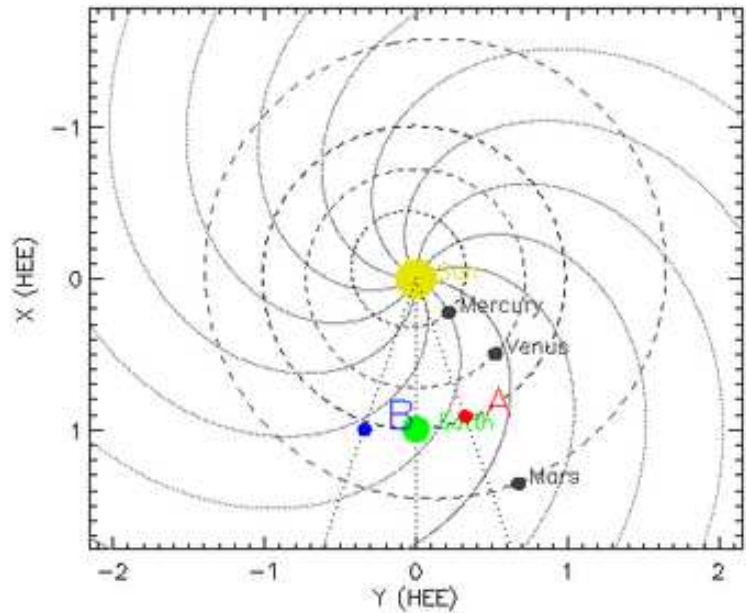}
 \caption{Positions of inner planets of the solar system and both STEREO spacecraft on November 21, 2007. Planetary orbits are shown as dahsed ellipses and continuous lines represent the nominal Parker spiral. Image obtained from the STEREO-Science Centre, https:\\stereo-ssc.nascom.nasa.gov/.}
 \label{fig:positions}
 \end{figure}

We can see in Figure~\ref{fig:positions}, obtained from the STEREO Science Center (https://stereo-ssc.nascom.nasa.gov), that STA and Mars were practically aligned, while the difference in heliocentric longitude between STA and Earth was $\sim$20$^\circ$ and that between STA and STB was $\sim$40$^\circ$. We can thus conclude that the same structure that hit STA, STB and ACE also reached Mars at later times.

\begin{figure}[h]
\centering
\includegraphics[width=1.0\textwidth]{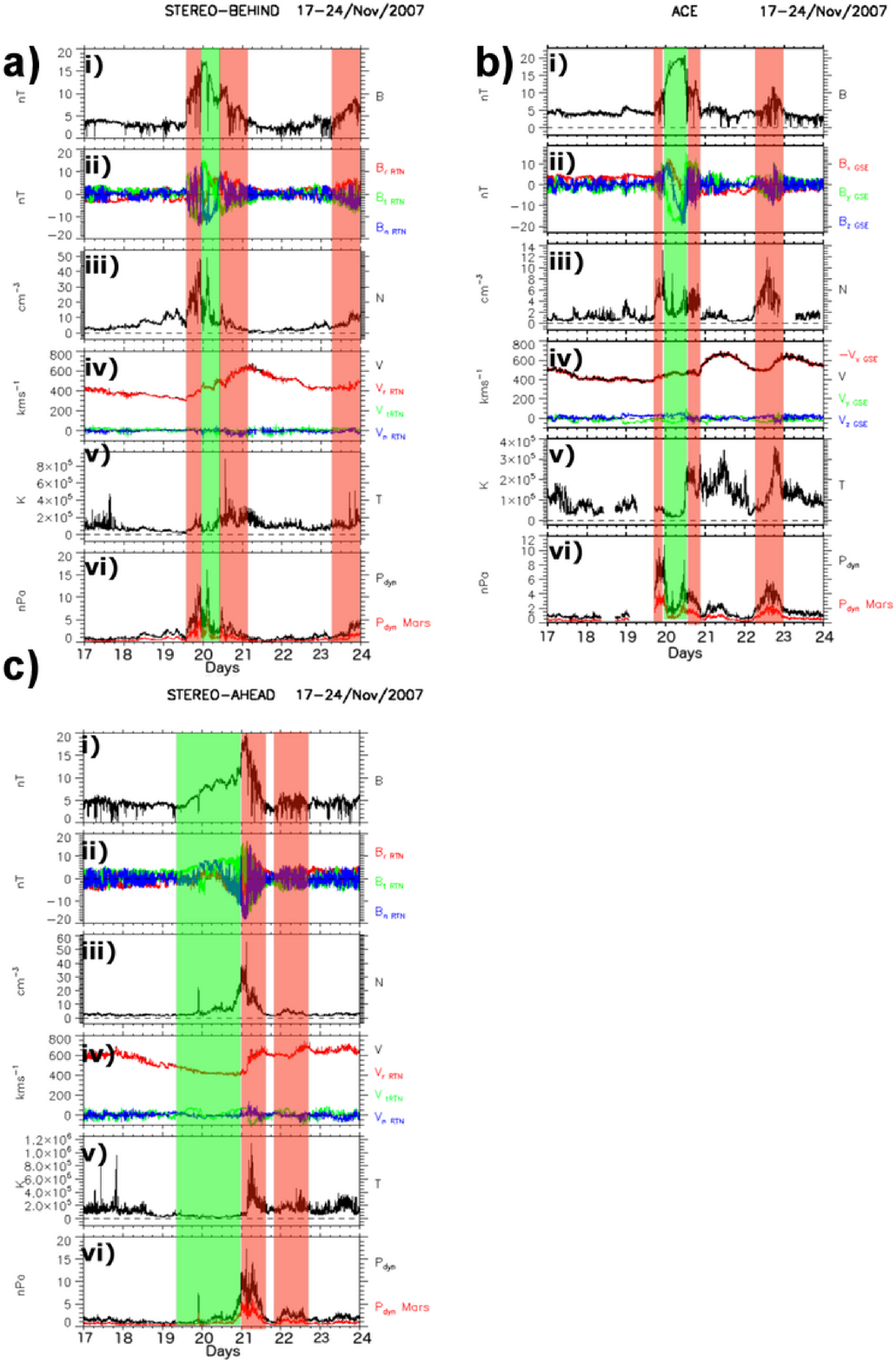}
\caption{Observations by Stereo B (a), ACE (b) and Stereo B (c). From top to bottom: IMF magnitude, IMF components, SW density, velocity (magnitude and components), temperature and SW dynamic pressure, P$_{dyn, SW}$. ICME is shadowed in green and the two SIRs in red.}
 \label{fig:imfplasma}
 \end{figure}

A detailed look at the STEREO and ACE data provides a more complete information on the SW transient itself. Figure~\ref{fig:imfplasma}a) shows data from STB. Panels exhibit: i) IMF magnitude in units of nT, ii) IMF components (nT) in RTN coordinates (Radial-Tangential-Normal coordinates: X axis points from Sun towards the spacecraft, while the Y axis is the cross product of the solar rotational axis and X, and lies in the solar equatorial plane), iii) plasma ion density (cm$^{-3}$), iv) SW velocity and components (kms$^{-1}$) in RTN system, v) temperature (K) and vi) SW dynamic pressure (P$_{dyn}$, nPa) at 1~AU (black) and scaled (as $1/r^2$) to Mars (red) between 17-24 November. We can see that on 19 November a SW structure arrived at STB. Its main signatures are increased B-field, density and P$_{dyn}$. During this transit, the SW velocity increases from 300 to 700~kms$^{-1}$. All this shows that the structure is a SIR (shadowed in red around a green shading). However smooth rotations of the B-field components on 20 November point to an ICME (shadowed in green) embedded inside the SIR, so the transient is a complex event. Towards the end of 23 November, there appears to be another structure in the data, which we classified as another SIR (shadowed in red).

Figures~\ref{fig:imfplasma}b) and c) show ACE and STA data, respectively, in the same format as Figure~\ref{fig:imfplasma}a). We can see that the appearance of the transients changes. At ACE the smooth B-field rotations, indicative of the ICME (green), appear closer to the upstream edge of the first SIR (red), while in STA data the first SIR (red) is preceded by the ICME (green).
The reason for such relative timings of the ICME and the SIR is because the top SW speed inside the SIR is larger than inside the ICME. At STB position the SIR had more time to expand outwards and had reached larger heliocentric distances than at STA. Thus, at STB the SIR already overtook the ICME, while at STA it had had barely caught up with it. This implies that we should expect Mars to be reached first by the SIR and later by the ICME. The second, less prominent SIR observed by the STB, was also observed by ACE and STA on 22 November.

Another SW feature can be observed from the Figures~\ref{fig:imfplasma}a)-c): the SW P$_{dyn}$ (panels vi) at 1~AU just before the arrival of the ICME+SIR structure is very low, about 0.81~nPa. If we scale it to the Mars distance at the time (1.53~A.U.), assuming that it varies with the heliocentric distance as $1/r^2$, we get even lower P$_{dyn}$ estimate of 0.35~nPa.

\section{Measurements at Mars}

\subsection{Overview}
\begin{figure}[h]
 \centering
 \includegraphics[width=0.8\textwidth]{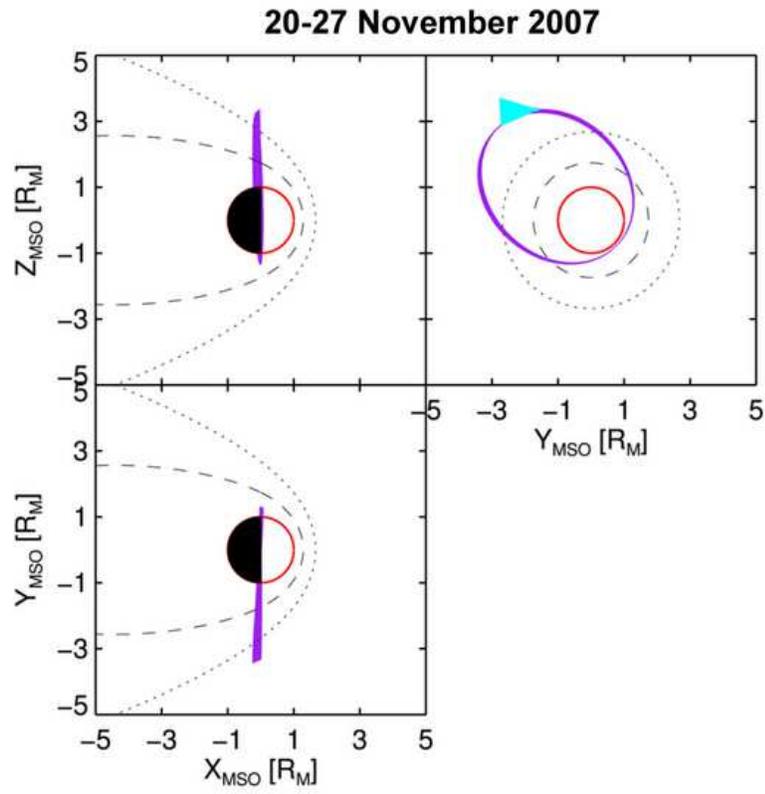}
 \caption{The orbit of Mars-Express in MSO coordinate system.}
 \label{fig:orbit}
 \end{figure}

Figure~\ref{fig:orbit} shows the MEX orbit during 20-27 November 2007.  We use the Mars-centered Solar Orbital (MSO) coordinate system in which the $x$-axis points from Mars to the Sun, the $y$-axis points antiparallel to Mars' orbital velocity while the $z$-axis completes the right-handed coordinate system. The coordinates are expressed in units of Mars radius (R$_M$). The dotted and dashed lines represent the BS and IMB according to the model by \cite{vignes:2000} and the cyan arrow shows the travel direction of MEX.

We can see that during the period of interest the MEX orbit was in the terminator plane. This enables us to estimate the trans-terminator ion flow much more directly. Its orbital period was 6 hours and 43 minutes. Its periapsis was on the southern hemisphere on the dusk side at the altitude of $\sim$300~km, while apoapsis occured at the distance of $\sim$3~R$_M$.

\begin{figure}[h]
 \centering
 \includegraphics[width=1.1\textwidth]{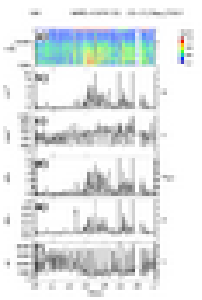}
 \caption{Mars-Express IMA observations during 20-27 November 2007. Panels show ion a) PDF spectrogram, b) ion density, b) velocity, c) thermal pressure e) dynamic pressure and f) temperature. The blue, red and green lines above the plots indicate quiet, SIR-perturbed and ICME-perturbed times, respectively.}
 \label{fig:mexglobal}
 \end{figure}

We first look at the MEX IMA data during 20-27 November 2007 (Figure~\ref{fig:mexglobal}). Panels exhibit: a) particle energy flux spectra (in units of keV/(cm$^2$ s ster keV)), b) plasma number density (cm$^{-3}$), c) velocity (kms$^{-1}$), d) thermal pressure (nPa) e) SW dynamic pressure (nPa) and f) temperature (eV). These data are obtained directly from the ClWeb plataform (http://clweb.irap.omp.eu/).

Strong variations of plasma parameters occur during each orbit because MEX enters very different plasma regions (SW, magnetosheath, ionosphere) with very different plasma properties. The largest densities are measured inside the ionosphere, while velocity, temperature and P$_{dyn,SW}$ peak when the spacecraft is either in the magnetosheath or in the SW. However we can see that these parameters also vary from orbit to orbit.

It can be easily seen that three events hit Mars. Their signatures are the increases of ion densities, thermal pressures and the ion dynamic pressure during times when MEX was in the SW. Increments of the SW velocity are not as clear as in the other parameters. The temperatures are lower during perturbed times than during quiet times. The blue, red and green lines above the plots indicate quiet, SIR-perturbed and ICME-perturbed times, respectively. The first event was observed at Mars from 22 November $\sim$9:30~UT - 24 November $\sim$9:00~UT (orbits \#4995-5002). The onset of the second event was on 24 November at $\sim$22:40~UT and it was observed until 25 November 14:30~UT (\#5004-5007).  The third event arrived at Mars on 26 November and it was observed only during two orbits (\#5009-5010) (see also Table~\ref{tab:times}). These times are  approximate since MEX performed observations only during $\sim$3~hr intervals centered around the times of its closest approach to the planet.

The P$_{dyn,SW}$ increases strongly during the first SIR (from $\lesssim$1~nPa to $\sim$7~nPa), which represents a 600~\% increase. During the second and the third event (the ICME and another SIR, respectively) the P$_{dyn,SW}$ increases to $\sim$5.5~nPa and $\sim$5~nPa, respectively, representing a 450~\% and 400~\% increases with respect to the quiet times before the arrival of the first structure.

Figure~\ref{fig:CR_HEND} shows fluxes of ions with energies between $\sim$30~keV and $\sim$195~keV measured by HEND detector on-board Mars Odyssey.  We can see that these fluxes were gradually increasing until the early 22 November at which point they decreased abruptly. After that there is one more spike at the beginning of 23 November followed by a drop. We can see from Figure~\ref{fig:mexglobal}b) that these dates coincide with the passage of the first SIR. Data in Figure~\ref{fig:CR_HEND} therefore indicate a Forbush decrease of energetic ions. Such decreases are commonly observed at Mars (e.g. \cite{mostl:2015, witasse:2017, guo:2018}).
We also examined the behaviour of ion fluxes with energies between 195~keV and 1~MeV (not shown) which did not exhibit such variations.

\subsection{Electron spectra}

\begin{figure}[h]
\centering
\includegraphics[width=1.2\textwidth, angle=0]{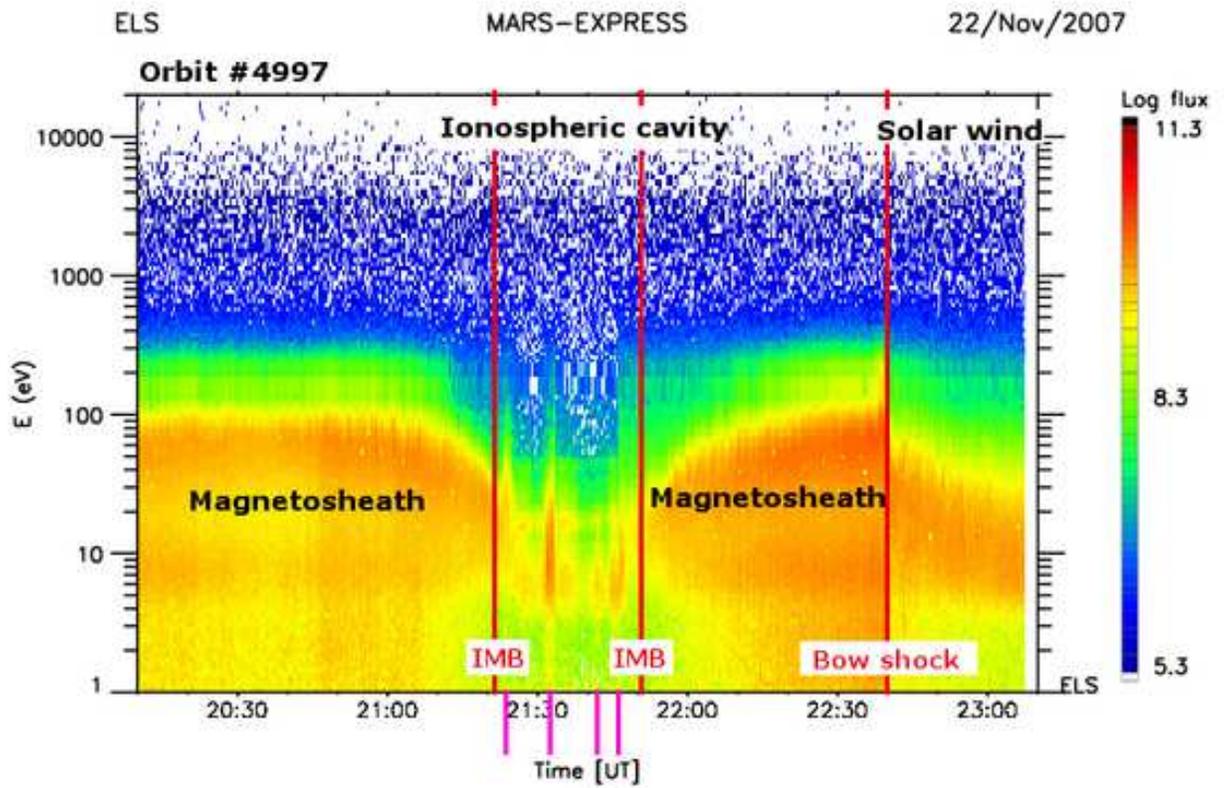}
\caption{ELS spectrogram during a portion of the orbit \#4997. Purple lines on the bottom mark the times when hot electrons appear inside the ionosphere.}
\label{fig:heated}
\end{figure}

Figure~\ref{fig:heated} shows spectra during the orbit \#4997 on 22 November 2007. The colors represent the logarithm of particle flux (in units of s$^{-1}$cm$^{-2}$sr$^{-1}$keV$^{-1}$) and the vertical red lines mark the times of plasma boundaries crossings. In the beginning of this time interval MEX observed hot electron population typical of that in the Mars' magnetosheath. At around 21:10~UT such population becomes gradually colder with its maximum energy decreasing until $\sim$21:20~UT, when the spacecraft crosses the IMB. After that only electrons with E$\lesssim$50~eV are observed, typical of those in the martian ionosphere. We call the whole interval when such electrons are observed simultaneously with ion of planetary origin, the ionospheric cavity. Later, at $\sim$21:50~UT the spacecraft enters the magnetosheath again and it remains inside it until $\sim$23:40~UT when an abrupt change in electron properties occurs. This is the time when MEX crossed the bow shock and entered into the solar wind.

\begin{landscape}
  \begin{figure}[h]
 \centering
 \includegraphics[width=1.5\textwidth, angle=0]{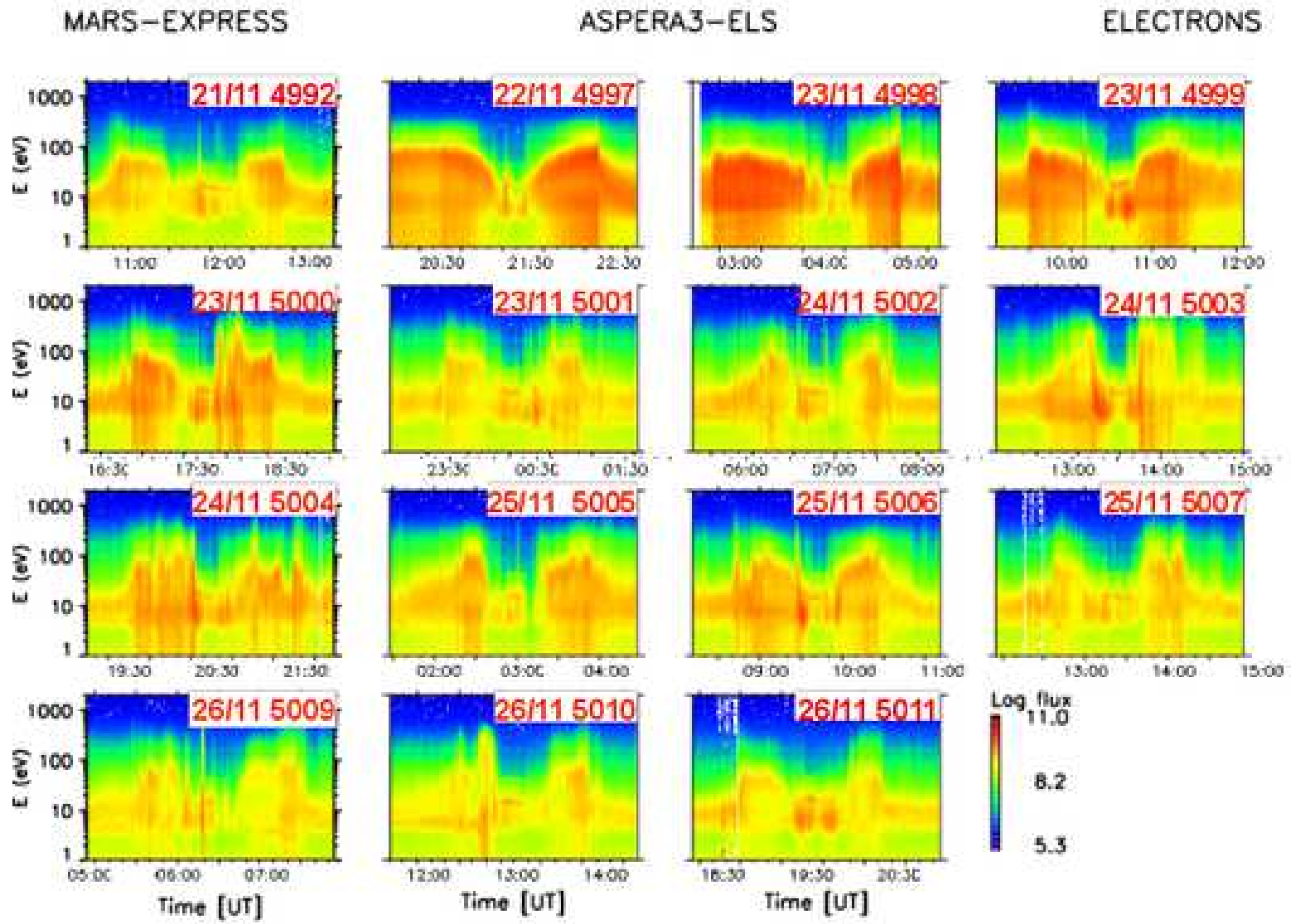}
 \caption{Electron spectrograms for successive orbits between November 21 and November 26 2007. The dates and MEX orbit numbers are provided for each panel.}
 \label{fig:els}
 \end{figure}
  \end{landscape}

Figure~\ref{fig:els} exhibits several ELS electron spectra during the period between November 21-26. Dates and orbit numbers are provided on each panel. The first panel (orbit \#4992) shows electron spectrum on a quiet day before the arrival of the SIRs and ICME.

The data for this orbit start at 10:26~UT on November 21. Relatively cold electron population with energies $\lesssim$30~eV indicates that the spacecraft was initially in the SW. After $\sim$10:42~UT the spacecraft observed an onset of intense fluxes (red and green traces) of hot electrons with energies up to $\sim$300~eV. This is indicative of the Martian BS crossing and the entrance in the Mars' magnetosheath. At $\sim$11:27~UT the spacecraft crosses the IMB and enters into the MPR and later into the ionosphere which is characterized by cold electrons with energies $\lesssim$50~eV. The spacecraft stays inside the magnetospheric cavity until $\sim$12:20~UT when it crosses the IMB again and enters the magnetosheath. At $\sim$12:56~UT the spacecraft crosses the BS again and enters the SW. During its time in the ionosphere MEX detects hot electrons between $\sim$11:50-11:53~UT with energies up to $\sim$400~eV. We note that during this orbit, both observed plasma boundary (BS, IMB) crossings are well defined and abrupt.

The next panel shows data during the orbit \#4997 which is when the first SIR already arrived at Mars (there are no electron data for orbits \#4993-4996). The most striking difference with the previous orbit is that the magnetosheath electron fluxes became much more intense, pointing to higher density in the magnetosheath. During its inbound flight MEX did not observe typical signatures of neither the SW nor the BS. This means that the BS moved away from the planet and was either crossed before the beginning of the ELS measurements or that it was located beyond the MEX orbit. The outbound BS crossing is easily identified at $\sim$22:40~UT. The electron fluxes in the SW are more intense than they were during the orbit \#4992.
Finally, the time that MEX spends inside the IMB is much shorter ($\sim$22 minutes) than before ($\sim$113 minutes). Inside it MEX encounters hot magnetosheath-like electron fluxes several times (see also the Figure~\ref{fig:heated}). These encounters may mean that hot magnetosheath plasma temporarily displaced ionospheric plasma at some locations crossed by the spacecraft, similar to what was reported by \cite{sanchezcano:2017}.

Intense electron fluxes are present throughout the following orbit \#4998. The main difference with the previous orbit is that the BS is observed during inbound crossing at $\sim$02:55~UT  and that the spacecraft remains inside the magnetospheric cavity for $\sim$30 minutes. Several hot electron fluxes are observed during ionosphere crossing possibly showing that parcels of hot magnetosheath plasma displaced ionospheric plasma at spacecraft altitudes. During the orbits \#4999--5001 the electron fluxes diminish and the magnetospheric cavity expands. During the orbit \#5002 the plasma configuration resembles that during the orbit \#4992. During orbits \#5003-5006 the electron fluxes increase again which indicates the compression of the planet's magnetosheath and induced magnetosphere. This time period coincides with the arrival of the ICME at Mars.

Electron fluxes during the orbit \#5003 exhibit several distinct features. The most obvious is that the intensity of the magnetosheath electron flux fluctuates, which is especially true during the outbound crossing.
During the inbound portion of the orbit, the electrons in the magnetosheath seem to heat more gradually and their maximum energy also augments gradually from the BS towards the IMB. Once inside the ionosphere, we see intense fluxes of electrons with very low energies between 5~eV and 10~eV. These fluxes are observed also during the outbound crossing from the ionosphere into the magnetosheath. The magnetospheric cavity is observed for $\sim$30 minutes in the data, which is less than during the quiet time orbit \#4992, but similar to the duration during the orbit \#4998 when the SIR was passing the planet.

The following orbits (\#5007-5008) correspond to intervals when the Mars' ionosphere/magnetosheath system was recovering. During the orbit \#5010 the electron fluxes increase again somewhat, which is a consequence of the arrival of the second, smaller SIR.

\subsection{Ion spectra}

\begin{landscape}
\begin{figure}[h]
\centering
\includegraphics[width=1.5\textwidth, angle=0]{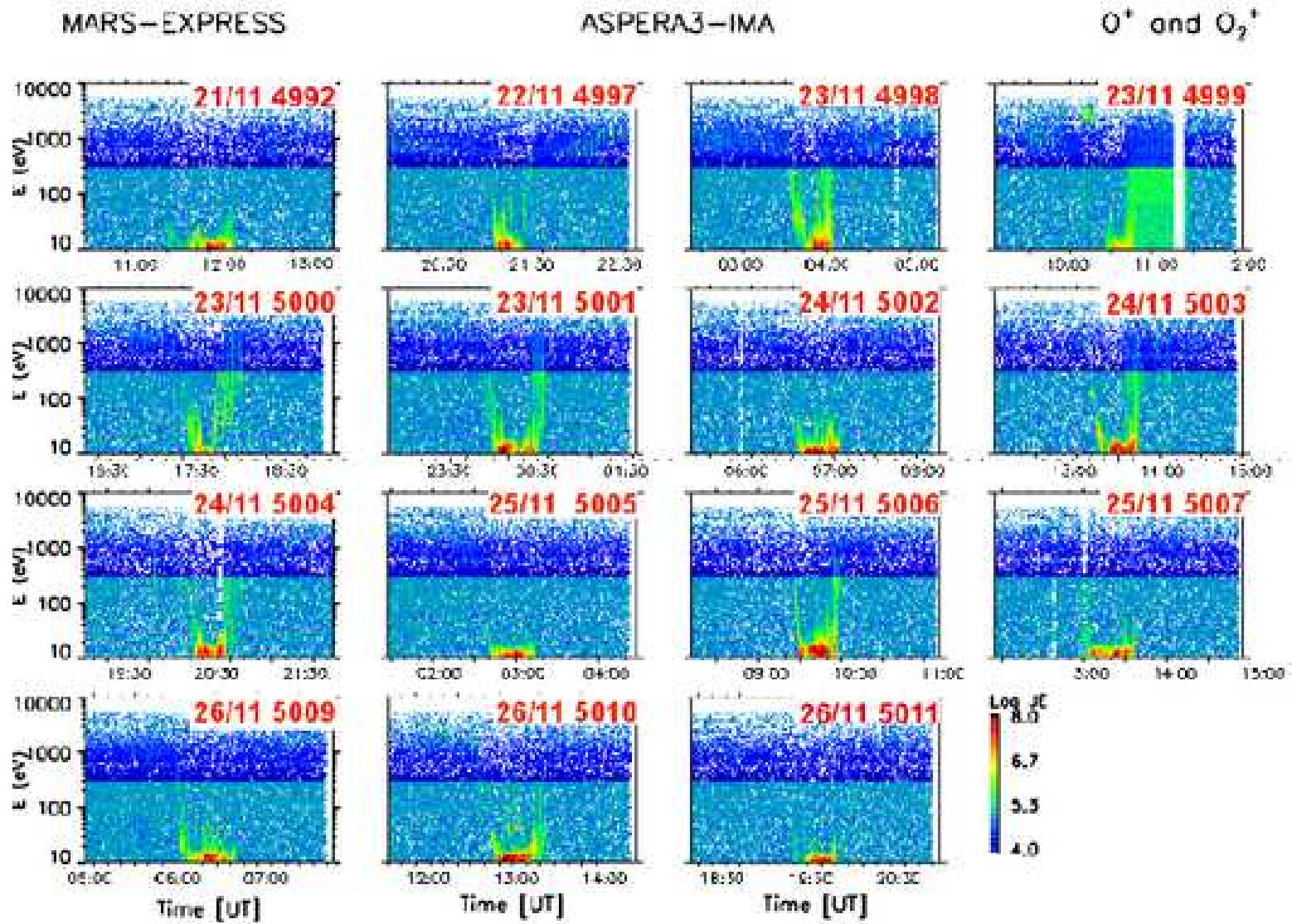}
\caption{Spectrograms of O$^+$ and O$_2^+$ ions during successive orbits between November 21 and November 26 2007.}
\label{fig:heavyions}
\end{figure}
\end{landscape}

Figure~\ref{fig:heavyions} shows spectra of O$^+$ and O$_2^+$ during the same orbits as shown in Figure~\ref{fig:els}.
We note that the wide trace at highest energies ($>$500~eV) that appears during the orbit \#4997-4998 is mainly due to contamination by magnetosheath protons, although the electrons with E$>$1~MeV and protons with E$>$20~MeV, such as those occurring during SEP/EPS events, may also generate IMA background counts during perturbed times (\cite{ramstad:2018}).

The spectra in Figure~\ref{fig:heavyions} show some features that were already observed in electron data, namely that the transition between the magnetosheath and ionosphere is abrupt and well defined during quiet times (orbit \#4992) and thus corresponds to the IMB. During perturbed times (orbit \#4997-5001, 5003-5006 and 5010) the ionospheric and magnetosheath proton fluxes are enhanced and the ionospheric cavity is reduced. During quiet times the heavy ions exhibit energies below 30~eV, but this changes during perturbed times when we observe accelerated heavy ion fluxes with energies up to $\sim$350~eV.

\subsection{Plasma boundaries}

\begin{figure}[h]
\centering
\includegraphics[width=1.1\textwidth, angle=0]{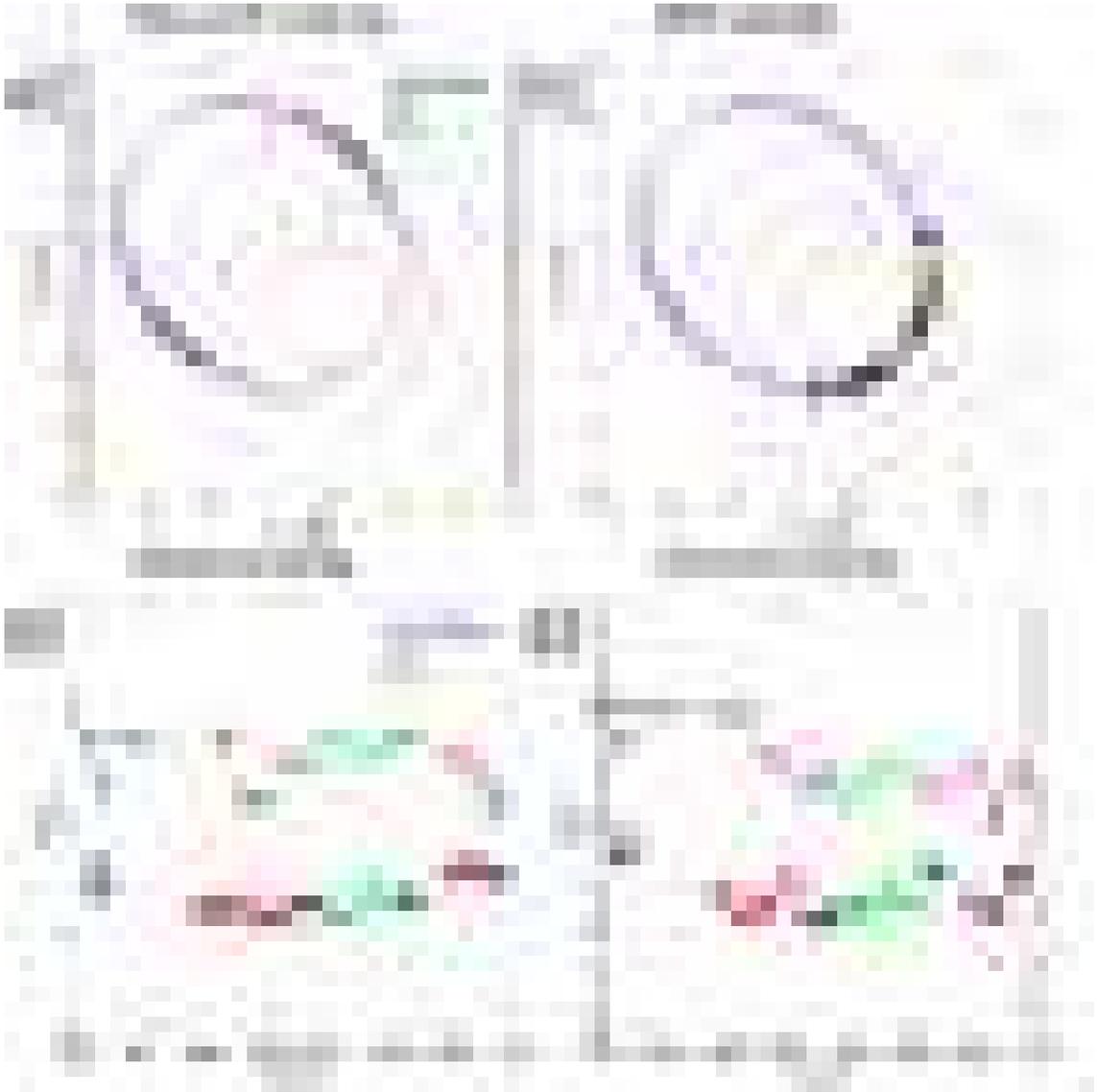}
\caption{Locations of bow shock crossings (left) and IMB crossings (right) during 20-27 November 2007 shown along the MEX orbit (top) and as a function of the MEX orbit number (bottom).}
\label{fig:boundaries}
\end{figure}

Figure~\ref{fig:boundaries} shows the behaviour of the plasma boundaries (bow shock, IMB) during 20-27 November 2007 determined from the electron spectra. Panels a) and b) show locations at which the first inbound and the last outbound crossings of the BS and the IMB, respectively, occurred in the Y$_{MSO}$-Z$_{MSO}$ plane. Blue, red and green symbols represent crossings during quiet times, the SIRs and the ICME, respectively. The dashed and dotted circles show the cross-section of the nominal BS and IMB, respectively, according to the \cite{vignes:2000} model.

Panels c) and d) show the distance from the center of Mars at which BS and IMB crossings were observed during inbound and outbound, respectively, parts of the orbits, as a function of MEX orbit number.

We can see that during times when the first SIR was passing the planet the BS location fluctuated most, between 2.2~R$_M$ (inbound crossing) and $\sim$3~R$_M$ (outbound crossing). During the orbit \#4997 the inbound BS crossing was not observed, so the BS was probably located even farther out from the planet.

This means that either the bow shock was located beyond where MEX began gathering data during this orbit.
During the ICME passage and the following quiet time orbits BS and IMB locations fluctuated much less. During the ICME passage the inbound (outbound) crossings occured at the distance of $\sim$2.7~R$_M$ ($\sim$2.4~R$_M$). After the ICME the inbound BS crossings occurred at monotonically diminishing distances between $\sim$2.7~R$_M$ and $\sim$2.2~R$_M$. The outbound BS crossings occurred at distances that fluctuated between 2.5~R$_M$ and 2.0~R$_M$.

The distances at which the IMB was crossed show a more systematic behaviour. They tend to be largest during quiet times (between 1.2 and 1.6~R$_M$ during inbound crossings and 1.1 and 1.7~R$_M$ during outbound crossings) with the exception of the inbound crossings during the second SIR.

The IMB was observed at smallest distances from the planet during the first SIR and the ICME passage. This shows that the magnetospheric cavity was compressed during these times.

\subsection{Plasma flow}

In this section we show how the SIRs and the ICMEs affected the transterminator plasma flow. Due to the limited availability of the data we limit the study of transterminator flow to O$^+$ and O$_2^+$ ion species and to radial distances below 2~R$_M$.

\begin{figure}[h]
\centering
\includegraphics[width=0.75\textheight, angle=0]{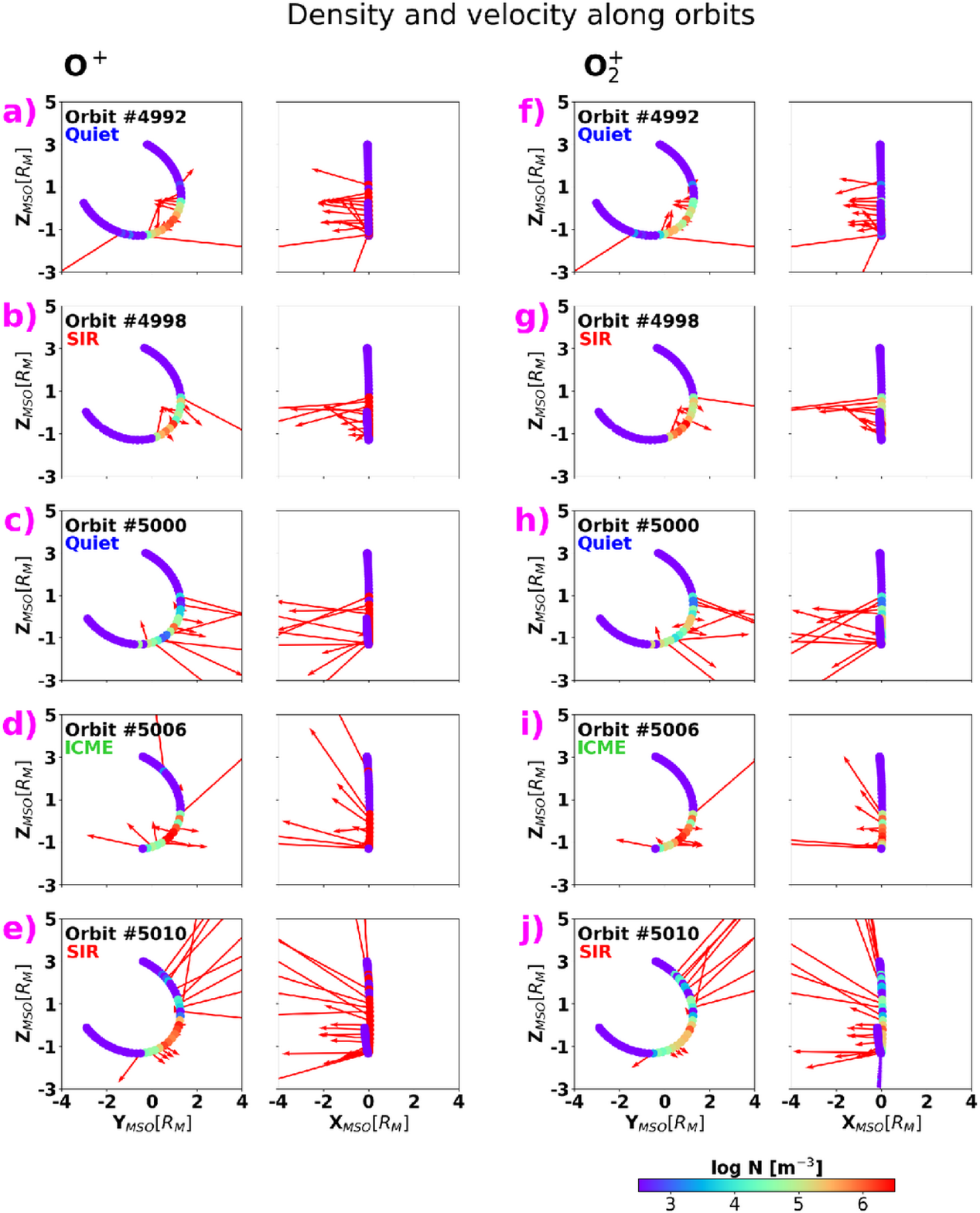}
\caption{O$^+$ and O$_2^+$ velocities (red arrows) and densities (color coded) along selected MEX orbits in the terminator and meridional planes.}
\label{fig:densvel}
\end{figure}

Figure~\ref{fig:densvel} shows how the ion density and velocity changed during selected orbits.
It exhibits data during six orbits during quiet times (panels a, c, f, h), during the SIRs passage (panels b, e, g, j) and during the ICME passage (panels d, i) in terminator and meridional planess.

The colors along the MEX orbit in Figure~\ref{fig:densvel} represent the logarithm of density in units of m$^{-3}$, while the arrows represent the projected plasma bulk velocity. The ion densities in the magnetosheath are much lower than in the ionosphere and are mostly represented with purple colors. In the ionosphere they are represented with green, yellow and red.

We can see that the velocity vectors of O$^+$ and O$_2^+$ show similar behaviours. They point in approximately the same directions although the O$^+$ vectors are longer (higher velocities) than those of O$_2^+$.

During the quiet-time orbit \#4992 most velocity vectors of either species have negative V$_{Y, MSO}$ and positive V$_{Z, MSO}$ components implying upward and dawnward movement. The plasma velocities near the apoapsis are very small with negative V$_{Y, MSO}$. The dominant velocity component is the negative V$_{X, MSO}$.

As the first SIR arrives (orbit \#4998) the velocities do not increase substantially, however they obtain a negative V$_{Z, MSO}$ component.
Velocities exhibit increased magnitudes during the quiet-time orbit \#5000. During the ICME-perturbed orbit \#5006 the velocity magnitudes decrease. What is interesting is that during the passage of the second SIR (orbit \#5010) the velocities of the ionospheric ions are rather low but those in the magnetosheath are strongly augmented. The magnetosheath velocities exhibit a strong positive V$_{Z, MSO}$ components which contrast the ionospheric velocities with negative V$_{Z, MSO}$.

\subsubsection{Transterminator flow}
The fact that the MEX orbit is practically in the terminator plane makes it possible to directly estimate the total transterminator flow of different ion species.

For this we first calculate the ion flow density j$_i$ perpendicular to the terminator plane at a time t$_i$ by multiplying the ion velocity along X$_{MSO}$ with number density at each MEX position. This was done for inbound parts of the orbits and only for distances $\lesssim$2~R$_M$ from the center of the planet:

\begin{equation}
 j_{i} = n_{i}V_{iX, MSO}.
\end{equation}

\begin{figure}[h]
\centering
\includegraphics[width=0.55\textheight, angle=0]{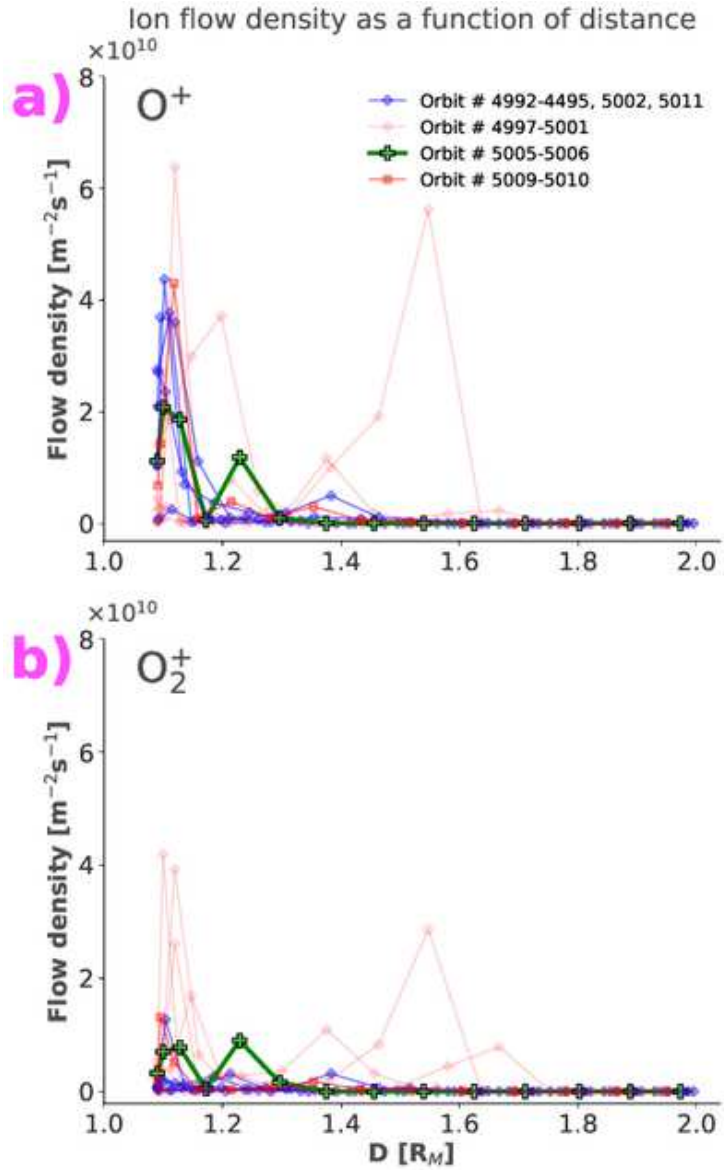}
\caption{O$^+$ and O$_2^+$ flow densities as a function of distance and orbit. The quiet-time values are plotted in blue while the values during the SIR and ICME-perturbed times are in red and green, respectively.}
\label{fig:flowdens}
\end{figure}

The results are shown in Figure~\ref{fig:flowdens}. O$^+$ values are plotted on panel a) and O$_2^+$ values on panel b). Blue, red and green traces and symbols denote quiet, SIR-perturbed and ICME-perturbed times, respectively. The values for the first SIR are presented as pale red traces and symbols so to distinguish them from the values during the passage of the second SIR. 

We can see that the behavior of the flow density is similar for both species but that O$^+$ flow densities are higher. This is expected since the O$+$ is the dominant species at radial distances such as those studied here (e.g. \cite{lundin:2009}). During quiet times the flow density diminishes with radial distance almost monotonically and is practically negligible for distances $\gtrsim$1.4~R$_M$. During perturbed times we observe peaks of flow densities at larger radii. In the case of the ICME this occurs at $\sim$1.25~R$_M$, while in the case of the first SIR there are large peaks at distances up to 1.7~R$_M$.

In the next step we multiply this flow density with the area of the ring with a radius R$_i$ and width $\Delta$R$_i$, where the latter is calculated as

\begin{equation}
\Delta R_i = \frac{R_i-R_{i-1}}{2} + \frac{R_{i+1}-R_{i}}{2} = \frac{R_{i+1}-R_{i-1}}{2}.
\end{equation}

R$_i$, R$_{i-1}$ and R$_{i+1}$ are MEX distances at successive times t$_{i-1}$, t$_i$ and t$_{i+1}$. The area of the ring A$_i$ is thus calculated as

\begin{equation}
A_i = 2\pi R_i\Delta R_i.
\end{equation}

During perturbed times the large $j$ values at large radial distances contribute most to the total flow since at these distances also the corresponding ring areas are largest.

\begin{figure}[h]
\centering
\includegraphics[width=1.0\textwidth, angle=0]{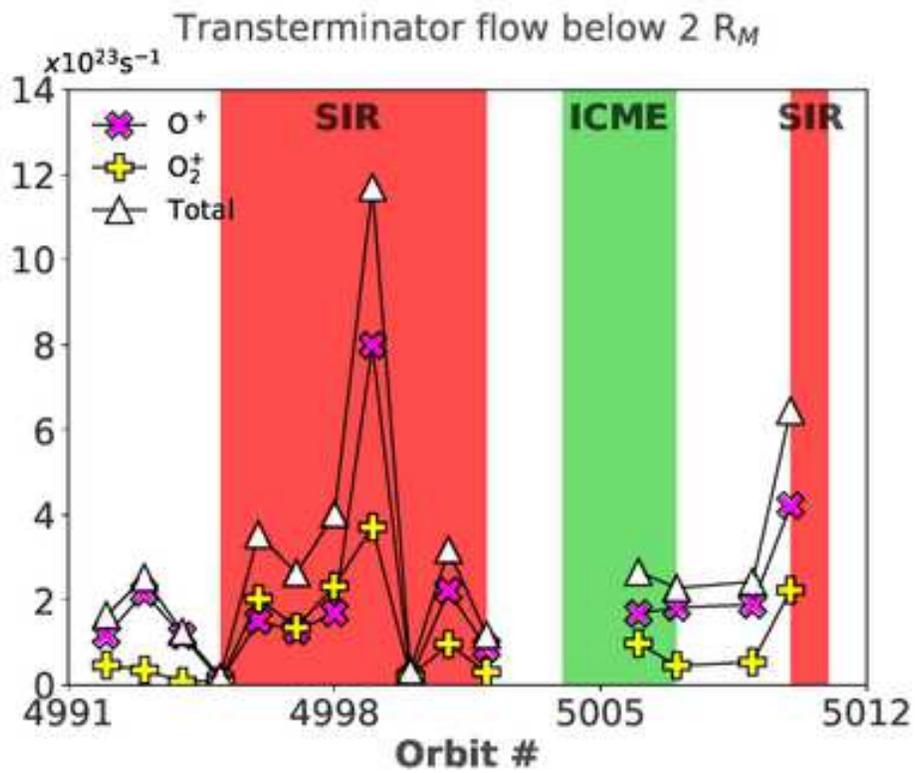}
\caption{O$^+$ and O$_2^+$ transterminator flow below 2~R$_M$ during the inbound portion of the orbit as a function of the MEX orbit number.}
\label{fig:transterminator}
\end{figure}

Finally, we sum the flow over all measurements during the inbound portion of the orbits. In the Figure~\ref{fig:transterminator} we show the ion  transterminator flow of O$^{+}$ and O$_2^+$. The O$^+$ values are represented with pink crosses, the O$_2^+$ values with yellow plus signs, while the total flow is presented with white triangles. SIR-perturbed and ICME-perturbed times are shaded in red and green, respectively.

The calculated flow varies greatly during perturbed times ranging between 8.3$\cdot$10$^{21}$-8$\cdot$10$^{23}$ particles per second. This is especially true for the flow during the passage of the first SIR when flows reach values up to a factor of $\sim$4 (O$^+$) and $\sim$8 (O$_2^+$) higher than those before the SIR arrival. During the ICME passage the flows are similar to those during quiet times, while they are enhanced again during the passage of the second SIR reaching values that are a factor of $\sim$1.9 (O$^+$) and $\sim$4.7 higher than those during quiet times.

\begin{figure}[h]
\centering
\includegraphics[width=1.0\textwidth, angle=0]{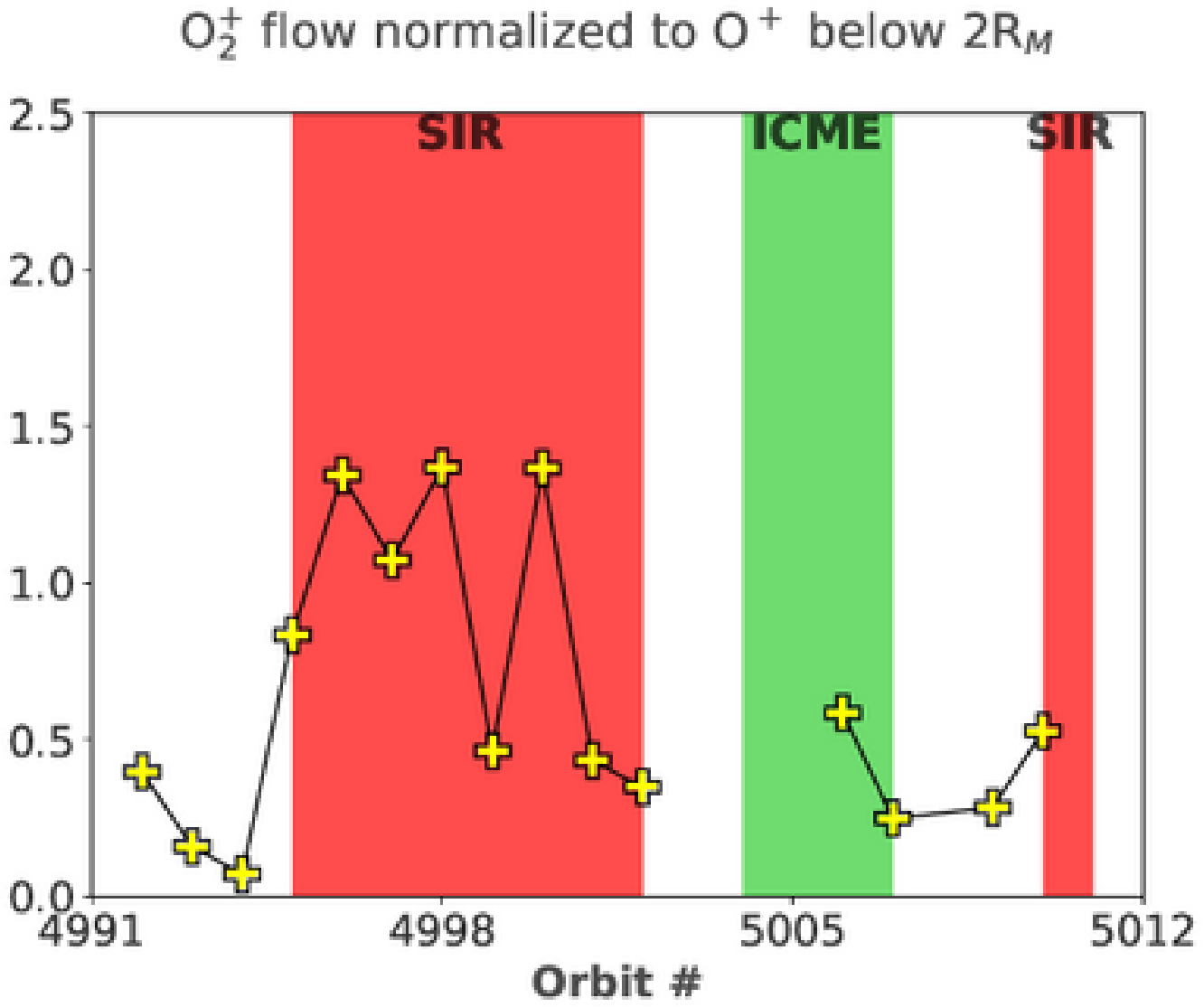}
\caption{O$_2^+$/O$^+$ flow ratio as a function of the MEX orbit number during the inbound portions of the orbit.}
\label{fig:ratio}
\end{figure}

Next we look at the Figure~\ref{fig:ratio} in which we show the O$_2^+$ flow normalized to that of O$^+$. This ratio varies between 0.07 and 0.4 during quiet times. The highest values of up to 1.37 are observed during orbits when the first SIR was interacting with Mars. The O$_2^+$/O$^+$ flow ratio during the ICME was 0.58 and that during the second SIR 0.53. This is somewhat higher than during quiet days.

\begin{figure}[h]
\centering
\includegraphics[width=1.0\textwidth, angle=0]{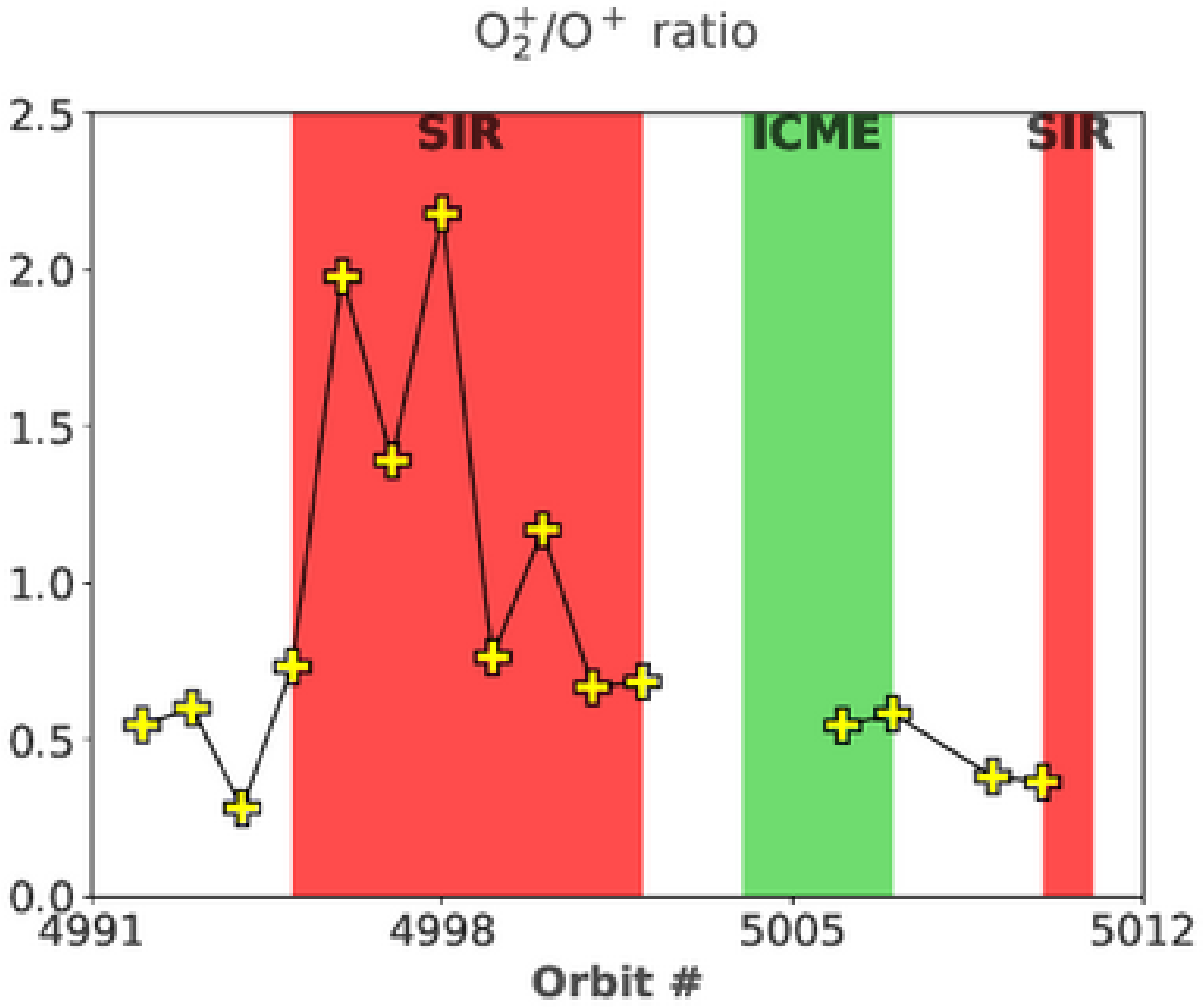}
\caption{O$_2^+$/O$^+$ ratio as a function of the MEX orbit number during the inbound portions of the orbit.}
\label{fig:nratio}
\end{figure}

Finally, Figure~\ref{fig:nratio} shows the density of O$_2^+$ normalized to that of O$^+$.
We calculate this ratio by following a similar procedure as for the flow calculation but without the multiplication by velocity. We can see that this ratio is $\lesssim$0.6 for all orbits except for orbits \#4995-5002 when the first SIR was passing the planet. During this time the O$_2^+$/O$^+$ density ratio reached the highest value of 2.2 during the orbit \#4998. It is interesting to note that during the times when ICME and the second SIR were passing the planet, this relative abundance was approximately the same as during quiet times.

\section{Discussion}

Here we compare the calculated BS and IMB distances and the transterminator flow properties to values from the existing literature.

We find that the BS location in the terminator plane varied by 0.8~R$_M$, from 2.2-3~R$_M$ (although during the inbound portion of orbit \#4997 the BS was not obserevd and was probably farther away from the planet) between 20-27 November 2007. This is more than the values reported by \cite{edberg:2010a} who found variations of 0.6~R$_M$ during two time intervals in 2005 and 2007. These authors came up with the equation relating the terminator BS distance R$_T$ and the SW magnetosonic Mach number (M$_{ms}$):
\begin{equation}
 R_T = -0.1M_{ms} + 3.3.
\end{equation}
From this we estimate the extreme M$_{ms}$ values at Mars during 20-27 November 2007 to vary beteen 11 and 3 during quiet and perturbed (first SIR) times, respectively. Both values could be considered as extreme according to the M$_{ms}$ distribution calculated by \cite{edberg:2010} which shows that the average M$_{ms}$ at Mars is 8.1 and almost all values lies between 5 and 11.

The IMB distance in the terminator plane varied between 1.1 and 1.6~R$_M$, so by 0.5~R$_M$ or 45~\%. The lowest values were found during the passage of the first SIR and the ICME. This coincides with the fact that these two structures caused the largest increase of the P$_{dyn, SW}$ and it also agrees with reports on BS/MPB compression during high solar wind pressure conditions (\cite{romanelli:2018}).

We should mention that in addition to P$_{dyn, SW}$ and the SW Mach number, there are other parameters that play an important role in the bow shock and IMB location variability. One of them is the solar EUV flux which may cause such variations on the order of days, years and through the solar cycle (e.g. \cite{hall:2016, hall:2019}). Short-term variations may be caused by solar flares (e.g. \cite{fletcher:2011}). However no flares were detected during the time period studied here. The second factor are crustal fields which were shown to have a small effect on the location of MPB (\cite{crider:2002, bertucci:2005}), but cause the north-south asymmetry of the bow shock (\cite{gruesbeck:2018}). The average IMB location variation on the southern hemisphere due to crustal fields has been shown by \cite{crider:2002} to be $\sim$200~km. However since the variations of the bow shock and IMB locations that we observe coincide with the arrival of the SW structures, we identify these structures to be the origin of the variations.

In contrast to our transterminator flow values it is more common in the literature to report escape rates. Escape flow and transterminator flow are of similar magnitude. It was was shown by, for example, \cite{ma:2004} that about half of the transterminator flow escapes.

\cite{nilsson:2011} estimated the heavy ion (O$^+$, O$_2^+$ and CO$_2^+$) net escape rate from Mars. The escape flux was calculated to be 2$\cdot$10$^{24}$~s$^{-1}$ which is similar to our highest estimate of transterminator flow of $\sim$1.2$\cdot$10$^{24}$~s$^{-1}$ during the SIR-perturbed orbit \#4999 and much higher than our lowest value of 3.4$\cdot$10$^{22}$~s$^{-1}$ during the quiet-time orbit \#4995.

\cite{franz:2010} estimated escape flows of O$^+$ and O$_2^+$ for several orbits, among them also for orbits \#5009 and \#5010. Their values are almost 2 orders of magnitude higher than our estimates. Part of the reason has to do with the fact that these authors model the ion distribution function at low energies. In our case however, the low energy ions are not taken into account. Due to this our densities may be underestimated and the same goes for flow densities and total flows.

\cite{edberg:2010a} estimated that during corotating interaction regions (a subset of SIRs) the heavy ion outflow from Mars increased by a factor of 2.5. Our transterminator flow values increase by a factor of $\sim$8 (O$_2^+$) and $\sim$4 (O$^+$), which is substantially higher.

\section{Conclusions}
In this work we study the interaction between SW structures (2 SIRs and one ICME) and Mars' plasma environment during 20-27 November 2007, when the Sun was in prolonged minimum of its activity. At the time the Sun, Mars and Earth were almost aligned, so we use ACE and STEREO probes in order to identify and characterize the SW structures that later hit the red planet. Mars was first hit by a SIR, then by an ICME and finally by a very weak SIR.
The MEX orbit at the time was in the terminator plane providing a unique view into the ion transterminator flow from the planet's ionosphere.

We show that the interaction between the three SW events and Mars' plasma environment resulted in the following features:
\begin{enumerate}
 \item We observe large displacements of Mars' IMB and the BS in the terminator plane. The IMB distance was reduced from 1.5~R$_M$ to 1.1~R$_M$, while the BS was located up to 3~R$_M$.
 \item The transterminator flow of O$^+$ and O$_2^+$ below 2~R$_M$ was enhanced during the passage of the two SIRs but not during the passage of the ICME. The largest flow was observed during the passage of the first SIR, when it was enhanced by a factor of $\sim$4 (O$^+$) and $\sim$8 (O$_2^+$) compared to quiet time conditions.
 \item The flow enhancement is in part due to large increase of ion flow densities at large radial distances (above $\sim$1.2~R$_M$). During quiet days the ion transterminator flow at these distances is practically negligible.
 \item The O$_2^+$/O$^+$ flow ratio below 2~R$_M$ varied greatly during perturbed time periods. It exceeded values of 0.5 during the ICME and the second SIR and 1.37 during the first SIR. During quiet times it was below 0.4.
 \item The O$_2^+$/O$^+$ density ratio was $\lesssim$0.6, except during the passage of the first SIR when it was above 1 with the maximum value reaching 2.2.
\end{enumerate}

\begin{figure}[h]
 \centering
 \includegraphics[width=0.8\textwidth]{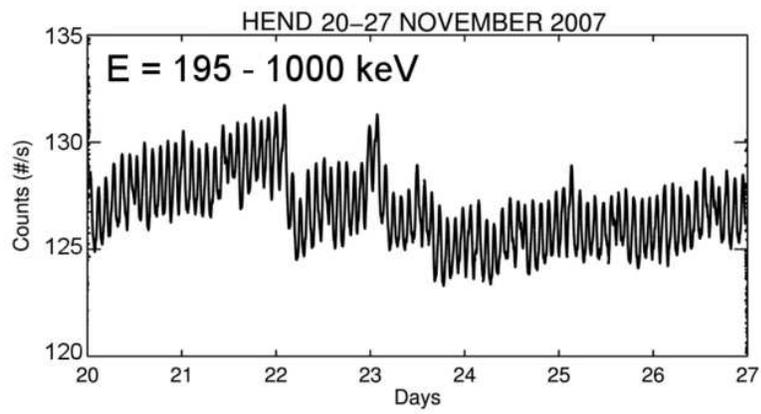}
 \caption{Suprathermal particles observed by HEND during 20-27 November 2007.}
 \label{fig:CR_HEND}
 \end{figure}

Hence we show that during solar minima the interaction of relatively moderate SW structures (SIR, ICME) with the Mars' plasma environment may result in large displacement of the IMB and the BS, in enhanced ion transterminator flow and the variations of its O$_2^+$/O$^+$ ratio. Especially strong was the interaction with the first SIR. This was the only structure intense enough so that it caused a Forbush decrease at Mars, as shown in the Figure~\ref{fig:CR_HEND}. During the passage of this structure, the P$_{dyn, SW}$ increased by $\sim$600~\% and it was much higher than during the quiet times and somewhat higher than during the passage of the ICME and the second SIR. This is consistent with \cite{curry:2015}, \cite{edberg:2009} and \cite{nilsson:2011} who showed that the ion escape increases with increasing P$_{dyn, SW}$. This points to the dynamic pressure probably being the most important parameter that governs the response of the Mars' plasma environment to SW perturbations.

\begin{table}[!h]
 \centering
 \begin{tabular}{l l l l }
 \hline
 \hline
 \multicolumn{4}{c}{\bf TIMES OF THE SW STRUCTURES AT DIFFERENT SPACECRAFT}\\
 \hline
 & SIR 1 & ICME & SIR 2\\
 \hline
 STA & 20 Nov 23:50 - 21 Nov 15:10 & 19 Nov 08:30 - 20 Nov 23:50 & 22 Nov 01:10 - 22 Nov 16:30 \\
 ACE & 19 Nov 18:00 - 20 Nov 22:40 & 19 Nov 23:30 - 20 Nov 12:30 & 22 Nov 06:00 - 23 Nov 23:30 \\
 STB & 19 Nov 14:30 - 21 Nov 06:00 & 19 Nov 23:30 - 20 Nov 15:50 & 23 Nov 16:50 - 24 Nov 07:30 \\
 MEX & 22 Nov 9:30 - 24 Nov 09:00 & 24 Nov 22:30 - 25 Nov 14:30 & 26 Nov 04:00 - 26 Nov 14:30 \\
  & Orbit \# 4996-5001 & 5003-5006 & 5009-5010\\
 \hline
 \end{tabular}
 \caption{Dates and times (UT) when STA, ACE, STB and MEX detected the two SIRs and the ICME.}
 \label{tab:times}
\end{table}

\typeout{}

\section*{Acknowledgements}
Authors acknowledge ClWeb (http://clweb.irap.omp.eu/), AMDA (http://amda.irap.omp.eu/) and ESA Planetary Science Archive Data (http://amda.irap.omp.eu/) teams for easy access and visualization of the data.  PK's work was supported by PAPIIT grants IA101118 and IN105620. ESA-ESTEC faculty funding is acknowledged. B.S.-C. acknowledges support through UK-STFC grant ST/S000429/1. This study was financed in part by the Coordena\c c\~ao de Aperfei\c coamento de Pessoal de N\' ivel Superior - Brasil (CAPES) - Finance Code 001.

\end{document}